# Making Palm Print Matching Mobile

Fang Li
School of computer engineering
Nanyang Technological University
Singapore
asfli@ntu.edu.sg

Maylor K.H. Leung
School of computer engineering
Nanyang Technological University
Singapore
asmkleung@ntu.edu.sg

Cheng Shao Chian
School of computer engineering
Nanyang Technological University
Singapore
Y060043@ntu.edu.sg

*Abstract*— With the growing importance of personal identification and authentication in today's highly advanced world where most business and personal tasks are being replaced by electronic means, the need for a technology that is able to uniquely identify an individual and has high fraud resistance see the rise of biometric technologies. Making biometric-based solution mobile is a promising trend. A new RST invariant square-based palm print ROI extraction method was successfully implemented and integrated into the current application suite. A new set of palm print image database captured using embedded cameras in mobile phone was created to test its robustness. Comparing to those extraction methods that are based on boundary tracking of the overall hand shape that has limitation of being unable to process palm print images that has one or more fingers closed, the system can now effectively handle the segmentation of palm print images with varying finger positioning. The high flexibility makes palm print matching mobile possible.

## I. INTRODUCTION

Personal identification and authentication have become a common task in today's highly advanced world where more and more day-to-day personal and business activities have been computerized [1-3]. Traditional identification and authentication systems rely on either a token item (e.g. a security pass card) or some knowledge only the user would know (e.g. passwords). Such systems are usually expensive in terms of time and resources to maintain and expand its usage. The most critical flaw of these systems is that since they do not use any inherent characteristics or attributes of the individual user, they are unable to differentiate between an authorized personnel and an impostor who have fraudulently come to possess the token or knowledge (such as stolen credit card or lost password). As such, these problems have led to system developers and researchers to explore into alternative solutions, and thus the intensified research arises on biometric identification and authentication systems [4-6].

Following this initial foray into biometric research, several forms of biometric systems based on different physiological or behavioral characteristics have been developed. The first commercial system, Identimat was developed in the 1970s [7]. The system was based on the measurement of the shape of the hand and the lengths of the fingers as the basis for personal identification. After that, various forms of biometric systems such as fingerprint-based systems and iris, retina, face, palm print, voice, handwriting and DNA technologies joined in over the years.

Among the leading biometric technologies, fingerprint-based system is the most prominent and widely used biometric technology, encompassing a market share of 58% in 2007 (A combine percentage of fingerprint and AFIS/Livescan technologies) [1]. The small size of the fingerprint-based device, ease of use and high accuracy has made it largely popular; however, as with most biometric solutions, it also has certain drawbacks. It is commonly found in most people that a layer of oil secretion or perspiration which emits from microscopic pores residing on the tiny ridges of the fingers will cover the surface of the fingerprint areas. As the resolution required for the fingerprint images are relatively high at approximately 500 dpi [7], this layer of secretion will render the fingerprint image capturing device useless or less effective in most cases. There are also cases whereby fingerprints wear away due to work or fraudulently scarred, all these will lower the effectiveness of fingerprint based systems.

In this project, we explore a relatively new biometric technology that employs palm print as the physiological characteristic that is used to differentiate between each unique individual. Palm prints are rich in features such as principal lines, wrinkles, ridge, datum points and minutiae points, all of which could be extracted at relative low resolution. Palm prints also have a much larger surface area as compared to fingerprints, which indicates that more features could be extracted from it, adding higher level of accuracy to it. These advantages place palm print-based technology as a promising biometric identification system.

Palm print recognition is an effective biometric technology that is gaining widespread acceptance and interest from researchers all over the world. As with most other biometric technologies, the process of palm print identification includes various stages from data acquisition, data pre-processing, feature extraction to matching process.

The main aim of this research is to improve the (Region Of Interest) ROI extraction process to increase the system robustness. By implementing and integrating a new square-based palm print ROI method into the previous application suite, the system is now able to overcome the limiting problem of failure to process palm print images with closed fingers, at the same time, to be (Rotation, Scaling and Translation) RST invariant, thus increase the flexibility of the system and in turn open up the possibility of bringing the palm print technology mobile. A new set of palm print image database captured using embedded cameras in mobile phone was created to find most robust ROI extraction techniques.

## II. OVERVIEW OF PALM PRINT MATCHING SYSTEM

Palm print recognition is an effective biometric technology that is gaining widespread acceptance and interest from researchers all over the world. As with most other biometric technologies, the process of palm print identification includes various stages from data acquisition, data pre-processing, feature extraction to matching process. The system overview is shown in Figure 1.

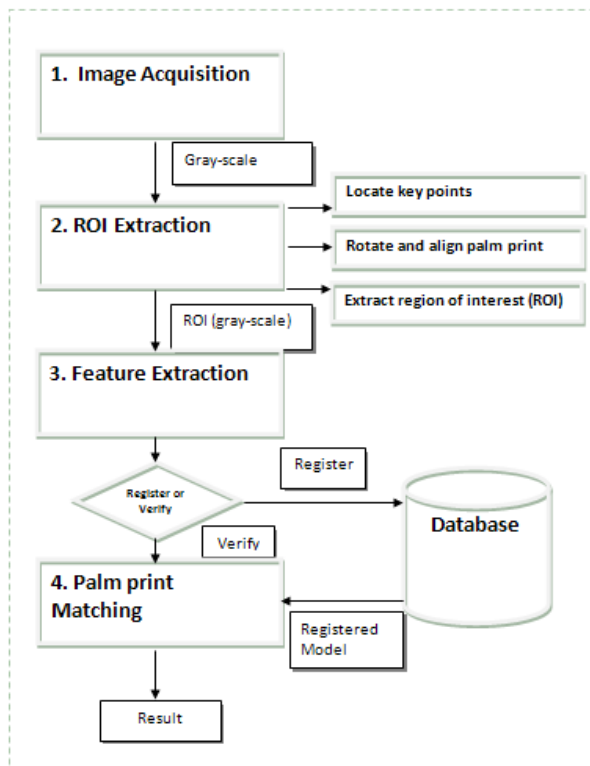

Figure 1. Palm print matching system overview.

## III. IMAGE ACQUISITION

In image acquisition stage, an image of the user's palm is captured by the system. Palm image can be acquired by a few methods. One method is to apply a uniform layer of ink on the palm and place the palm on a paper; the paper is then scanned into PC to obtain a digital image. This method is regarded as an "off-line" method.

Another method for palm image acquisition is to use digital devices to photograph the palm and immediately obtain a digital image stored in the system. Such digital device can be a digital camera [8], which is shown in Figure 2, or scanner [9], which is shown in Figure 3. This method is regarded as an "on-line" method.

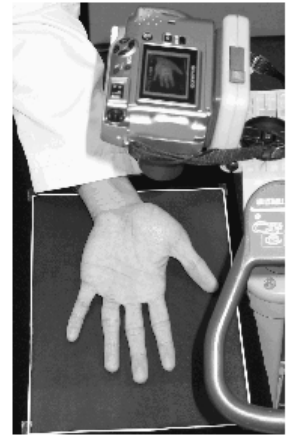

Figure 2. Acquisition of a typical image sample using digital camera [8].

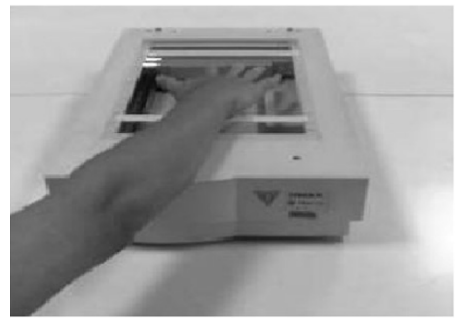

Figure 3. Acquisition of a typical image sample using scanner [9].

Inked palm print is not a good choice for mobile matching, so we only discuss online palm print in this research.

## IV. PRE-PROCESSING: REGION OF INTEREST (ROI) EXTRACTION

After image acquisition, the raw input is passed to the verification stages to perform various image processing operations. Normally, the raw image consists of palm, fingers, wrist, and a substantial amount of background area. For the verification process, only the inner area of the palm is of interest. The system needs to trim away those unwanted portions of the raw image to reduce the amount of computation required in the subsequent stages. Another problem with the raw input image is that the location and orientation of the palm is not fixed.

Palm prints may show certain degree of distortion as the image may be captured at different times and rotated at different angles. Furthermore, it could also be affected by varying conditions in terms of temperature, humidity and lighting condition. As such, even if two images are from the same palm, we could end up with a conclusion that they are from different sources.

Palm print preprocessing, the segmentation process, involving the correction of such distortion and placing all the palm prints in the database under the same coordinate system and orientation such that the proper expected area of each palm print can be extracted for use in accurate feature extraction and matching, greatly improves the efficiency and correctness of the identification system.

The output of palm print segmentation is a sub image, known as the Region of Interest (ROI) or central part sub-image of the palm print, which is cut out from the original input image. This sub image represents the inner area of the palm, where most of the palm print features are within this area. Those methods can be further classified into two different classes: square-based ROI extraction and inscribed circle-based ROI extraction [8]. As the circle-based approach consumes a significantly higher amount of computation resources, based on this experimental outset, we will only focus on square-based ROI extraction approach throughout this research.

The basic idea of square-based ROI extraction technique as demonstrated in Figure 4 is to determine key gaps-between-fingers point on a palm print, thereafter two selected key points are lined up to form the y-axis, subsequently, a second line, which is the x-axis is drawn perpendicular to the y-axis through the middle point to form the origin. Finally, a square with a fixed size, ROI, is extracted under this coordinate system. All the pixels within this ROI are retained for further processing whereas the area outside the window is ignored and discarded. The essential rule in this extraction process is that the portion of the image extracted should be available in all palm prints from the database and there are sufficient palm print features for extraction and comparison. Moreover, the extraction should be RST invariant and gesture difference tolerant.

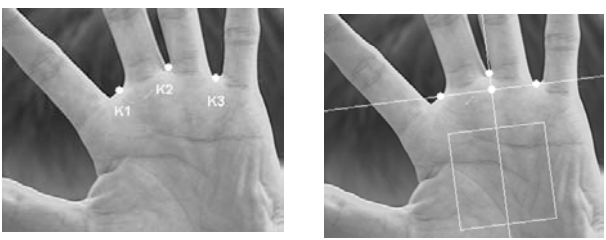

Figure 4. Basic idea of square-based segmentation technique.

All ROI extraction techniques principally rely on the determination of the key gaps-between-fingers to draw up the coordinate system through the use of boundary tracking algorithm. Thus this imposes a limitation in that for a palm print image to be properly segmented, the fingers in the image need to be sufficiently separated in order for the boundary tracing to work, and thus the key points are determined accurately.

Such requirement led to the development of image acquisition devices, which are shown in Figure 5, 6, and 7, that utilizes pegs [10-13] to restrict the movement and positioning of the hand during the acquisition process in order to improve the image quality, to solve RST problem, and to ensure that the fingers are properly separated.

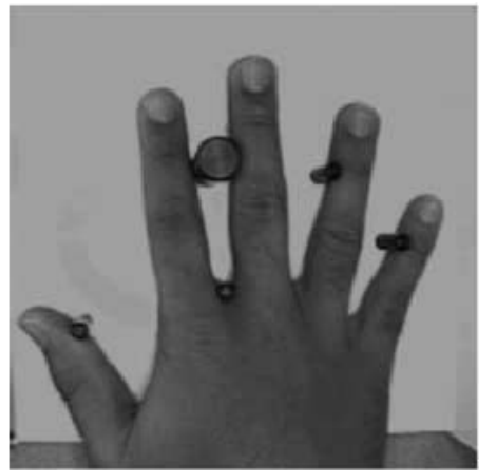

Figure 5. Palm image acquisition device with pegs [10].

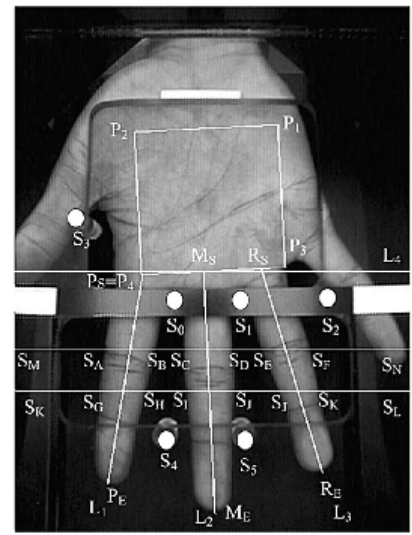

Figure 6. Palm image acquisition device with pegs and fixed ROI location [11].

PolyU-Online-Palm print-II is a benchmark palm print database for research purpose [5]. In the database, all the palm images are captured by a specially designed device [13] as shown in Figure 7. The distance from palm to camera, pegs between fingers, and the board between middle and ring fingers are fixed. A lot of researchers use this benchmark database and the device constraint ROI extraction process based on this database is shown in Figure 8 [12].

In this project, some built in cameras in mobile phones are used for image acquisition. The advantage of using those built in cameras as input devices is that the user does not need to purchase any device because of the popularity of mobile phones. Nowadays, almost everyone has a mobile phone with built in camera. Using these cameras, the palm print verification system can be easily integrated into any security systems without requiring extra image acquisition devices.

## V. IMPROVED SQUARE-BASED ROI EXTRACTION METHOD

The aim of this research is to implement an improved ROI extraction technique that is robust enough to overcome this reliance of a standard image acquisition device, and thus able to make use of the ubiquitous digital cameras and embedded cameras in mobile phones to perform the image capturing process, which in turn, will widen the scope of applications for palm print-based systems.

The improved square-based ROI extraction technique consists of the following steps:

Step 1: Gray image to binary image.

Step 2: Contour of hand generation

The above two step are standard ones in all palm print preprocessing system, so details are omitted here.

Step3: Straight Lines Extraction

As mentioned in section 4, the locations of three feature points need to be detected in order to set up a coordinate system for palm print alignment. These key points lie on the bottom of valleys between fingers. By observing the line pattern of the boundary image, the bottom of valley is a short curve joining the edges of adjacent fingers. The key points are best represented as the mid-points of those short curves.

To locate the mid-point, one method is to first find the line ($Lm$) that divides the inter-finger space into halves, the intersecting point between $Lm$ and the bottom curves of the valleys is one of the desired key points. Usually, the edges of two adjacent fingers form a V-shape. An angle can be established by extending the V-shape edges until they intersect. The line $Lm$ can be found by calculating the bisector of such angle. The method is illustrated in Figure 9.

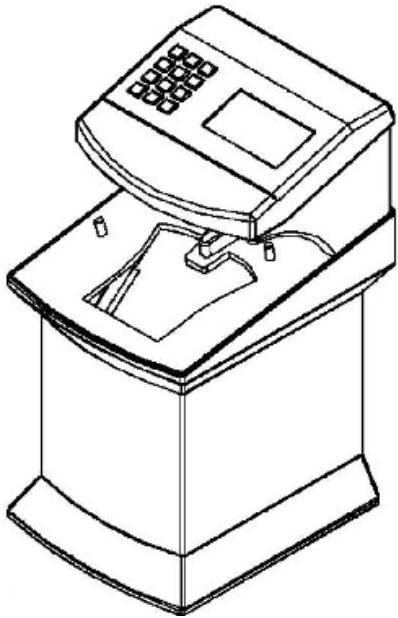

Figure 7. Patent Palm image acquisition device designed by D. Zhang et al [13].

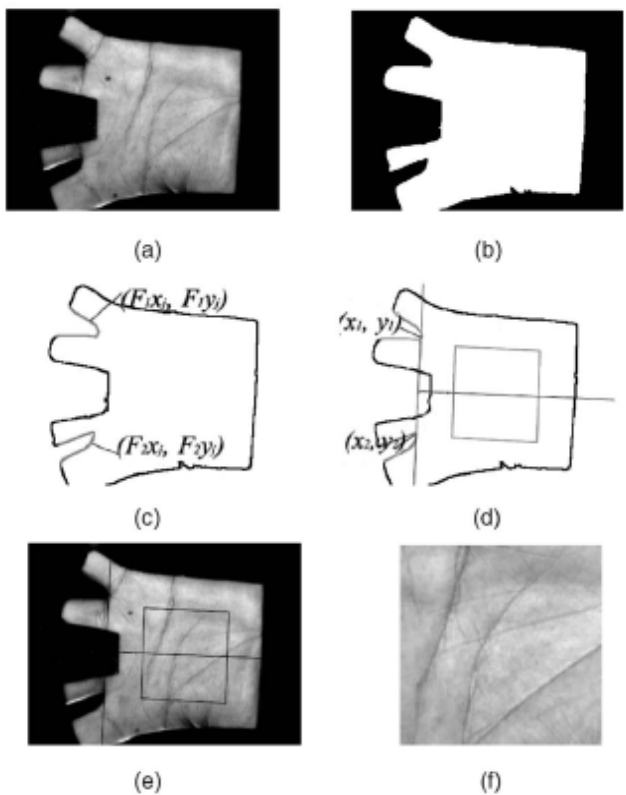

Figure 8. ROI extraction process based on database PolyU-Online-Palm print-II [12].

However, such devices are normally fixed to a site, and too bulky to move around for usage. This lack of mobility consequently results in the restrictive applications of palm print authentication system [14-16].

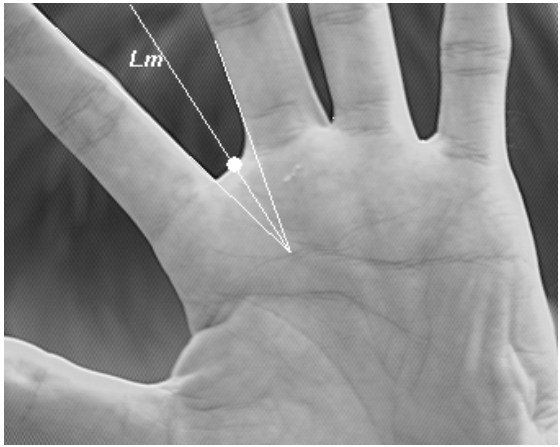

Figure 9. Method of Locating Key Point.

Generally, it is not easy to locate the V-shape edges directly. The problem can be solved by finding the parallel edges of each finger first, then for every two adjacent fingers, select an appropriate edge from each finger, and use the two selected edges to form a V-shape pair. This approach is feasible since detecting parallel line pairs are much easier than detecting V-shape line pairs directly.

Based on the method above, the task of locating the three key points in the boundary image can be divided into four steps:

Step 3.1: Straight lines extraction: Find the straight lines in the edge image, select the long ones which have the potential to be edges of fingers.

Step 3.2: Parallel lines grouping: Group the extracted long lines into parallel pairs, each pair represents two edges of a finger.

Step 3.3: V-Shape lines grouping: Reorder the parallel pairs and group the lines into V-shape pairs, each pair represents the edges of two adjacent fingers.

Step 3.4: Key points location: Form an angle for each V-shape pair and calculate the bisector of each angle. Find the intersecting points between the boundaries of inter-finger valleys and the calculated bisectors. The intersecting points represent the desired key points.

*Straight lines extraction*

The edge pixels of the binary image are traversed and the contour representation of the edge image is generated. Contour is a compact way to represent the shape of an image. The edge pixels are formed into separate groups, where each group represents a connected curve. The Dynamic Two-Strip (DYN2S) algorithm [17] is employed to perform the curve fitting operation. If a curve has only small variation (i.e. it is closed to a straight line), then this curve will be reduced to a single line segment which approximates to the original curve. After processing, the number of line segments in the image is greatly decreased, which reduces the difficulty of identifying the finger edges.

In order to get longest possible lines which have the high chance to represent the location of the fingers, broken line connection algorithm describe in [18] is adopted here to offset the possible broken lines issue. The details of the boundary of the palm are not necessary to analyze, so tiny straight lines are excluded.

*Parallel Lines Grouping*

Finding the parallel pairs in this method means finding the two lines which are the edge of the finger. To find the parallel lines, firstly one line must be taken and then check every other line whether it is the parallel partner of the first line or not. After that, the next line must be taken as the first line, and check the other line to find its parallel partner. It must be done until all of the lines have been checked whether they have parallel partner or not. The extracted parallel pairs extracted from the palm image in Figure 9 are shown in Figure 10.

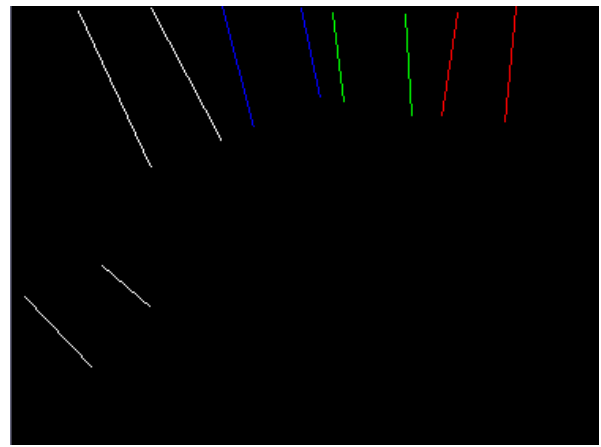

Figure 10. The parallel pairs.

*V-shaped Lines Grouping*

V-shaped lines grouping means detecting the two line in between two fingers. Firstly, the parallel pairs obtained from the previous algorithm must be sorted from the leftmost to the rightmost. Since the parallel pairs obtained from the previous algorithm are stored using the 2-D array, it is easy to sort the parallel lines and get the V-shaped line pairs. To get the V-shaped pairs, it is basically shifting all the lines by one in to the right as below:

- Sort the parallel line pairs, so that the line pairs are stored in left to right order.
- For each parallel pair Pi in the sorted array, form a V-shape pair with the right edge of $P_i$ and the left edge of $P_{i+1}$ (i = 0..I-2, where I is the total number of parallel pairs).

The result of this phase can be seen in the Figure 11. The same color lines identify the lines in the same pair:

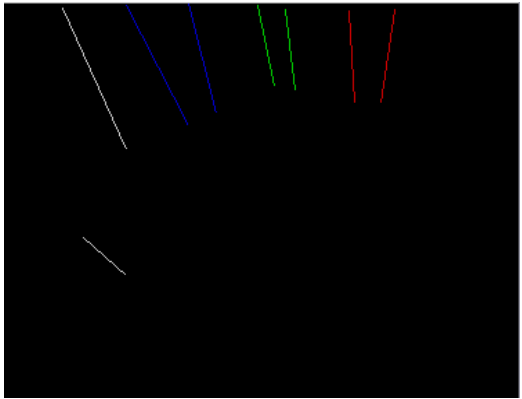

Figure 11. The V-shaped pairs.

*Key Points Location*

The location of the key points can be calculated with utilizing the V-shaped pairs obtained from the previous stage. The algorithm can be explained as below:

a) Extend the lines of the V-shaped pair until they intersect each other

b) Create the new center lines of an inter-finger space which has the bisector of angle formed by the two lines of the corresponding V-shape pair.

c) The key points for palm print alignment can be located by calculating the intersecting points between the bottom curves of valleys and the center lines of the corresponding inter-finger spaces.

However the above algorithm cannot work correctly in some images due to the exceptional cases in the v-shape pairs. As demonstrated in Figure 12, the exceptional cases occurred usually when the inter-space between the finger does not wide enough. Different person has different finger shape, when people close their figures tightly, the gap shape in-between two figures can be different. It can be either a parallel V-shape line pair with the same angles' value or an intersected V-shape pair intersected above the expected key point.

To overcome this issue, additional checking is added to check which category the pair belongs to. Once it is confirmed, proper action will be taken to set the center line in above step b) as follows:

–  If it is a parallel line pair, the center line is set to be the middle line of these two parallel lines.

–  If it is an intersected line pair, the center line is set to be the line which divides the intersected angle into two halves.

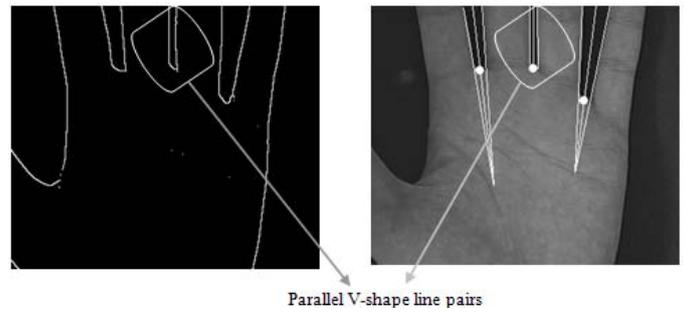

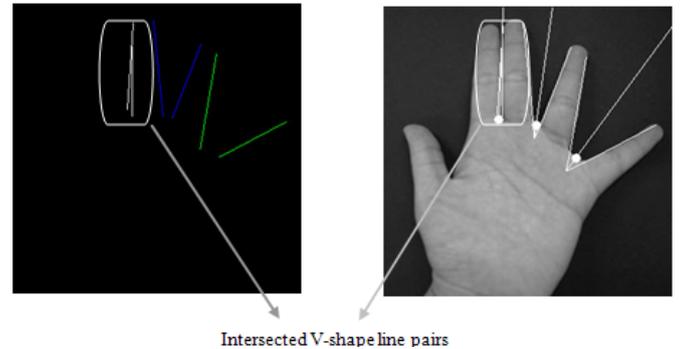

Figure 12. Exceptional cases occurred usually when the inter-space between the finger does not wide enough.

*Selecting Main Key points*

When the number of key points identified is more than 3, it means that the key point between thumb and index finger is also detected. The system must exclude it since the key points needed to locate the ROI are only K1, K2, and K3. It is necessary to select three desired key points out of the four. Among the four points, every three adjacent points can form a triangle. There are two such triangles in total. This is illustrated in Figure 13.

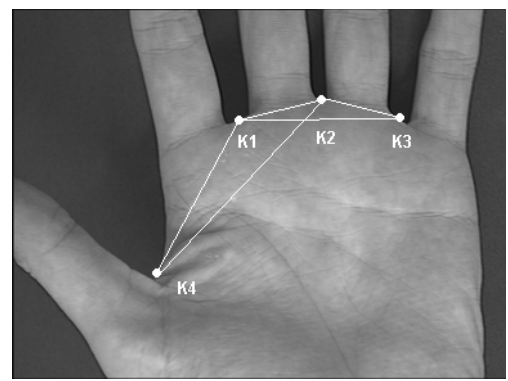

Figure 13. Triangles Formed by Four Key Points.

It is noticed that the triangle containing K4 has a greater height than the other triangle, because K4 is relatively far away from the other points. Hence, the desired key points can be determined by selecting the triangle with smaller height.

Step 4: Establish the ROI Window

After the desired key points (K1, K2, and K3) are located, the coordinate system can be established in order to create the ROI of the palm using the way described in section 4. The size of ROI is dynamically determined by the distance between K1 and K3. It makes the ROI extraction scale invariant. The distance between the camera and palm does not affect the range of the region we extract. It gives the user maximum freedom. A few samples of ROIs extracted from palms with flexible positions are shown in Figure 14. Despite the scale and gesture, the consistent regions are extracted.

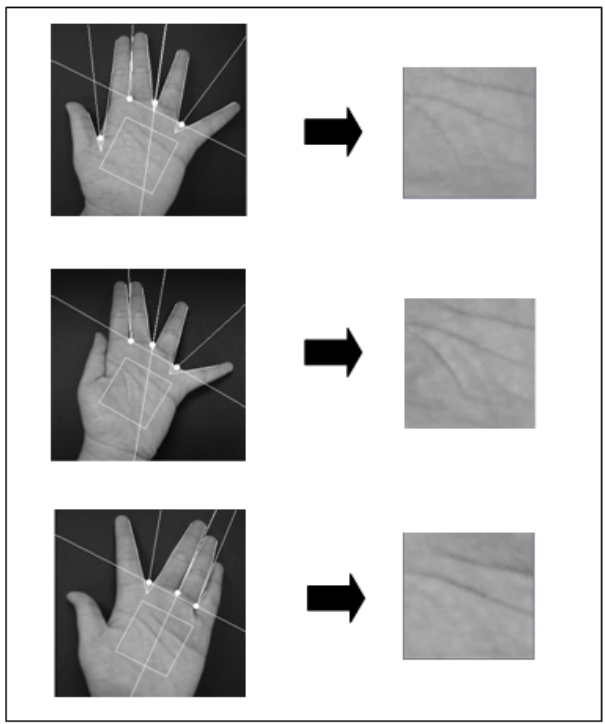

Figure 14. ROIs extracted with flexible palm positions.

## VI. LINE EXTRACTION

Principal lines and course wrinkles are more obvious than original ROI for human vision to judge the efficiency of ROI extraction. Line information is highlighted by following steps [19]:

i. Apply the averaging mask
ii. Applying the line detection masks
iii. Threshold the image
iv. Line thinning

Figure 15 illustrates the above steps through the first example in Figure 14. The line structure is highlighted that it is much easier for human vision to tell.

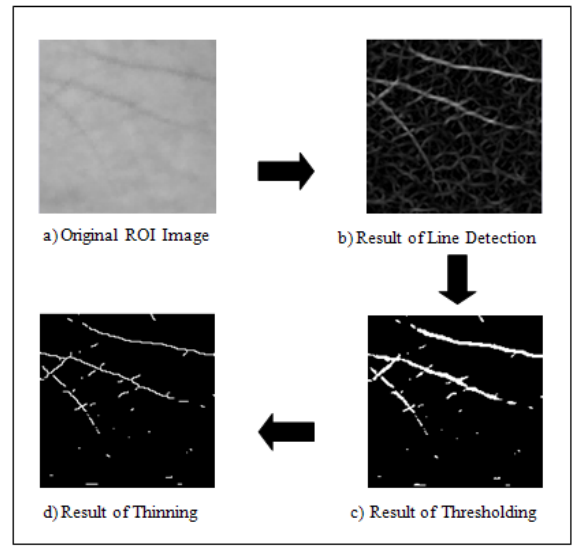

Figure 15. Sample of line extraction.

## VII. EXPERIMENT AND RESULTS

As the main aim of this research is to implement RST invariant ROI extraction method that is capable of handling palm print images that are varying in terms of fingers positioning and distance to camera. A new mobile palm print database with one thousand five hundreds photos is formed.

In this research, palm print images are captured using three mobile embedded cameras with different resolutions from two different mobile phones. During the image capturing process, no fixed pegs were used to restrict the movement, rotation and stretching of the hands. Each device is used to capture images of both hands from thirty subjects. Figure 16, 17, and 18 show the sample images captured by different mobile phones. With each hand, five photos are taken for each of the five different positioning of the hands in order to test the robustness of the algorithm.

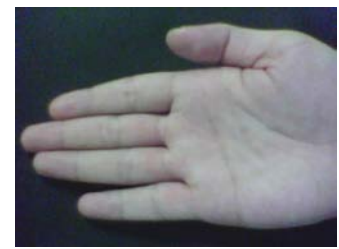

Figure 16. Palm print captured using D810 VGA camera.

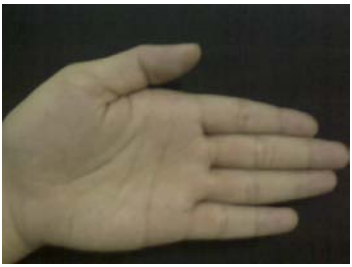

Figure 17. Palm print captured using D810 2MP camera.

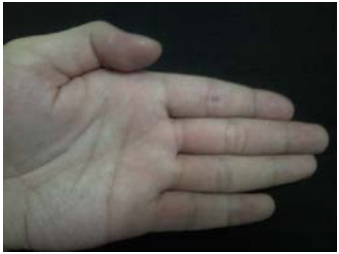

Figure 18. Palm print captured using SGH-i900 5MP camera.

The complete procedure to extract ROI in this research is illustrated by one example shown in Figure 19.

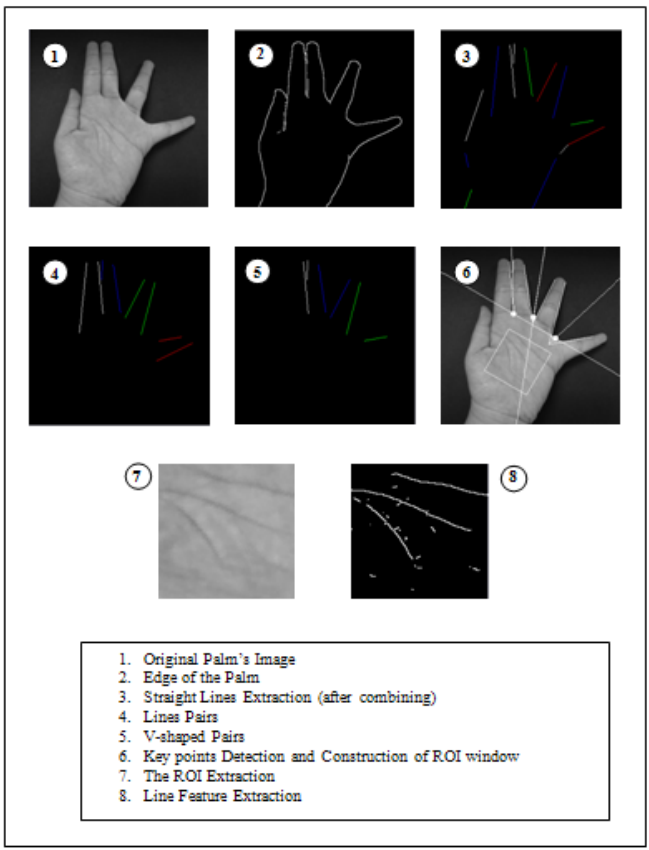

Figure 19. The complete procedure to extract ROI in this research.

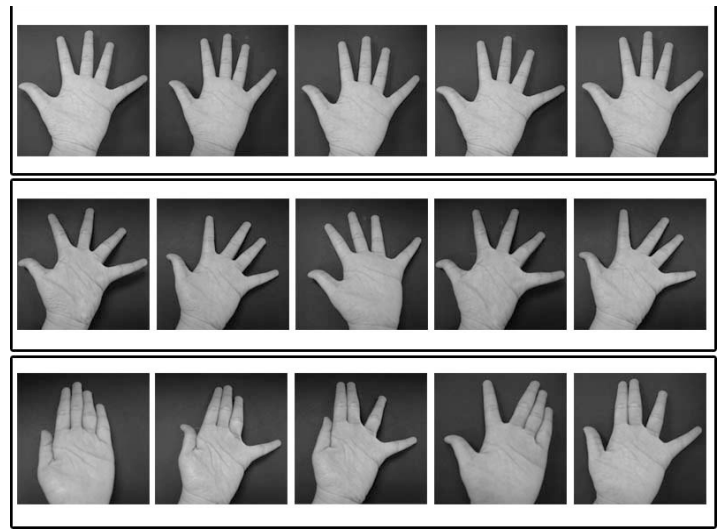

Figure 20. The images used for experiment with various positioning of fingers.

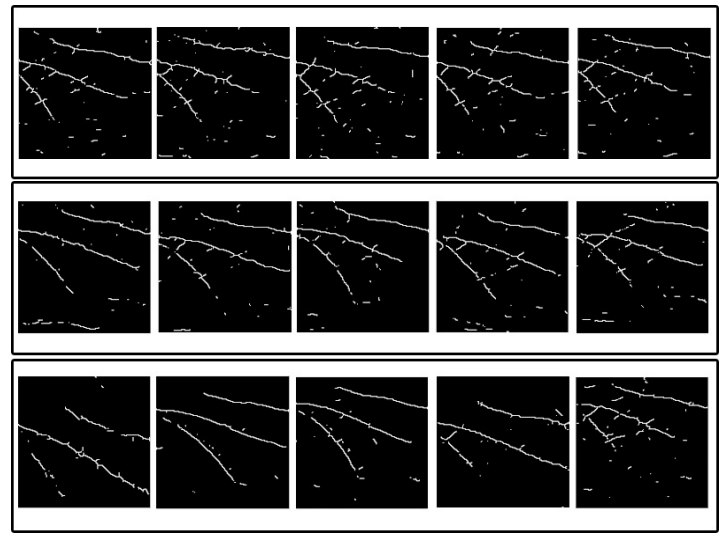

Figure 21. The extracted ROIs of palm prints in Figure 20.

To test the robustness of the new algorithm, 15 palm images from one person used in the experiment. There is no constraint on palm position and orientation. The subject can put her palm freely while taking the pictures. As for gesture, these palm images can be classified into 3 classes according to the abnormality's level of both the palm's shape and position. The first class consists of 5 normal-shaped palm images with optimal in-space distance between the fingers. The second class consists of 5 normal-shaped palm images with various positions and in-space distance between the fingers. The last class consists of 5 abnormal-shaped palm images with various position and in-space distance between the fingers.

Figure 20 shows the images used in the experiment. Images in the same row represent the same class according to the sequence of class. E.g., the images in the first row are classified in to first class, and so on.

The results of all images are expected to be approximately the same, since they are taken from the same person's palm.

Figure 21 shows the result of ROI extraction of the images in Figure 20. Observe from Figure 21, the new technique is approved to be successful as the ROIs are acceptably consistent.

Take image pair 11 and 14 as example. The images' scales, locations, and orientations vary a lot, the extract two ROIs are acceptably consistent. It shows that the new technique can tolerate RST variance well.

Among the three gesture categories, the ROIs in same row are very similar to each other. ROIs in second class are more similar to those in first class than third class. Although the results in third class are not perfect, they are still satisfactory since they are consistently representing the similar significant palm print features as other two classes.

## VIII. CONCLUSION

The improved square-based palm print ROI extraction method was successfully implemented and integrated into the current application suite. In comparison to other ROI extraction methods that are based on boundary tracking of the overall hand shape with limitation of being unable to process palm print images that have one or more fingers closed, the system can now effectively handle the segmentation of palm print images with varying finger positioning. It opens up the possibility of bringing the palm print technology mobile.

Through the experiments and findings in this research, it was found that the images captured with latest model of the mobile camera at five mega pixels, the verification rate was close and comparable to those of the palm print images in the research benchmark. Thus, with the certain continuous advancement in mobile camera technologies, it would be in no time that the usage of palm print authentication in mobile systems be proved as viable and practical for widespread applications.

However, researchers have proved that the identification performance of most of the unimodal methods are not satisfactory due to a variety of problems such as noise in data, restricted degrees of freedom, non-universality, and low accuracy. Fusing several modalities together might be able to make system more robust [20]. In future, we will investigate a feature to work together with palm print without compromising the key advantage of our current system, that is, high degree of freedom.